\documentclass[prl, twocolumn, superscriptaddress]{revtex4}

\usepackage{amsmath}  % need for subequations
\usepackage{amsfonts}
\usepackage{graphicx}   % need for figures
\usepackage{hyperref}   % use for hypertext links, including those to external documents and UrLs

\usepackage{color}
\newcommand{\V}[1]{\ensuremath{\boldsymbol{ #1}}}

\newcommand{\del}{\V{\nabla}}

\newcommand{\eqnref}[1]{eqn.~\ref{#1}}
\newcommand{\figref}[2][]{Figure~\ref{#2}\lowercase{{\bf#1}}}
\newcommand{\fref}[2][]{Figure~#2\lowercase{{\bf#1}}}
\newcommand{\figureref}[2][]{Figure~\ref{#2}\lowercase{{\bf#1}}}
\newcommand{\Figureref}[2][]{Figure~\ref{#2}\lowercase{{\bf#1}}}

\newcommand{\Tableref}[2][]{Table~\ref{#2}{{\bf#1}}}
\newcommand{\pp}[1]{\lowercase{{\bf#1}},}
\newcommand{\ra}{\hfill}

\begin{document}

\title{How superfluid vortex knots untie}
\author{Dustin Kleckner}
\affiliation{James Franck Institute and Department of Physics, The University of Chicago, Chicago, IL 60637, USA}

\author{Louis H.~Kauffman}
\affiliation{Department of Mathematics, Statistics and Computer Science, University of Illinois at Chicago, Chicago, IL, 60607, USA}

\author{William T.~M.~Irvine}
\affiliation{James Franck Institute and Department of Physics, The University of Chicago, Chicago, IL 60637, USA}

\begin{abstract}
Knotted and tangled structures frequently appear in physical fields, but so do mechanisms for untying them.
To understand how this untying works, 
we simulate the behavior of 1,458 superfluid vortex knots of varying complexity and scale in the Gross-Pitaevskii equation.
Without exception, we find that the knots untie efficiently and completely, and do so within a predictable time range.
We also observe that the centerline helicity -- a measure of knotting and writhing -- is partially preserved even as the knots untie.
Moreover, we find that the topological pathways of untying knots have simple descriptions in terms of minimal 2D knot diagrams, and tend to concentrate in states along specific maximally chiral pathways.
\end{abstract}

\maketitle

\begin{figure}[!b]
\includegraphics{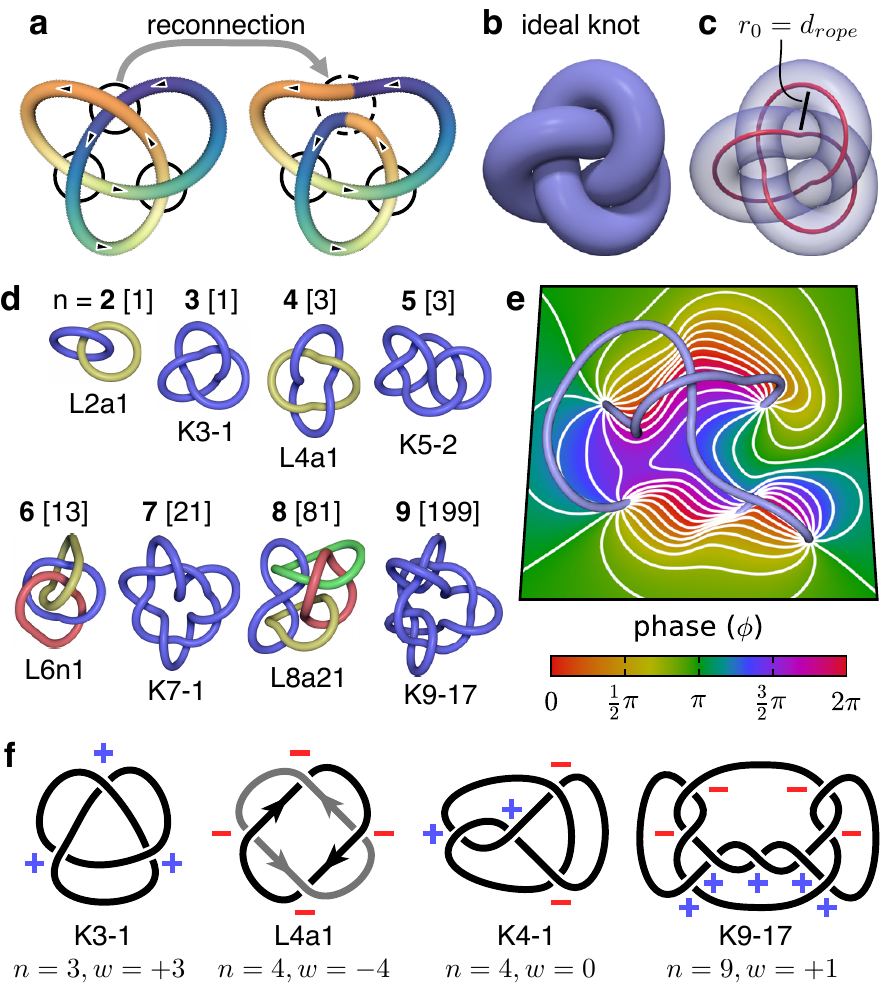}	
\caption{
	\label{fig-ideal}
	Reconnection events and vortex knots.
	\pp{A} A schematic of a vortex reconnection event, in this case converting a trefoil knot (K3-1) to a pair of linked rings (L2a1).
	\pp{B} An `ideal', or minimum rope-length, trefoil knot.
	\pp{C} Using the center-line of an ideal knot provides a consistent, uniform geometry for any knot or link; nearby strands are exactly spaced by the rope diameter, $d_{rope}$, which becomes the characteristic radius, $r_0$, of the loops which compose the knot.
	\pp{D} Example ideal configurations of topologies with different minimal crossing number, $n$.  
	The number of topologies excluding mirrored pairs is indicated in square brackets.
	\pp{E} A 2D slice of the phase field of a superfluid wavefunction with a knotted vortex line (light blue).
	\pp{F} Example minimal knot diagrams; in each case the topology cannot be represented by a simpler planar diagram.  The chirality of each crossing is indicated.
}
\end{figure}

Tying a knot has long been a metaphor for creating stability, and for good reason:
untangling even a common knotted string requires either scissors or a complicated series of moves.
This persistence has important consequences for filamentous physical structures like DNA, the behavior of which is altered by knots and links~\cite{Sumners1995,Wasserman1986}.
An analogous effect can be seen in physical fields, e.g., magnetic fields in plasmas or vortices in fluid flow; in both cases knots never untie in idealized models, giving rise to new conserved quantities~\cite{Moffatt1969, Woltjer1958}.
 At the same time, there are numerous examples in which forcing real (non-ideal) physical systems causes them to become knotted: 
  vortices in classical or superfluid turbulence~\cite{Moffatt1992a, Barenghi2007}, magnetic fields on the surface of the sun~\cite{Cirtain2013}, and defects in condensed matter phases~\cite{Tkalec2011}. 
This presents a conundrum: why doesn't  everything get stuck in a tangled web, much like headphone cords in a pocket~\cite{Raymer2007}? 

In all of these systems, `reconnection events' allow fields to untangle by cutting and splicing together nearby lines/structures (\figref[A]{fig-ideal})~\cite{Proment2012, Kleckner2013, Cirtain2013, Wasserman1986,Tkalec2011,Bewley2008}.
As a result, the balance of knottedness, and its fundamental role as a constraint on the evolution of physical systems, depends critically on understanding if and how these mechanisms cause knots to untie.

Previous studies of the evolution of knotted fields have been restricted to relatively simple topologies or idealized dynamics \cite{Kleckner2013,Dennis2010,Martinez2014,Wasserman1986}.
Here, we report on a systematic study of the behavior of all prime topologies up to nine crossings by simulating isolated quantum vortex knots  in the Gross-Pitaevskii equation (GPE, \eqnref{eqn-gpe}).
We observe that all knots untie, regardless of topological complexity or scale, and do so along preferred topological pathways.
As the vortices untie, the loss in topology is compensated by a gain in the `coiling' of the unknotted vortices. 
Crucially, these results are determined by the geometry of the vortex knots, rather than the details of the vortex reconnections.

The quantum counterpart of smoke rings in air, vortices in superfluids or superconductors are  line-like  phase defects in the quantum wavefunction, $\psi(\V x) = \sqrt{\rho(\V x)} e^{i \phi(\V x)}$, where $\rho$ and $\phi$ are the spatially varying density and phase (\figref[E]{fig-ideal}).
This quantum wavefunction can be mapped to a classical fluid velocity and density via the Madelung transform: $\V u = \del \phi$; $\rho = |\psi|^2$~\cite{Madelung1927}.
A simple description of the time evolution of this superfluid wavefunction is given by the Gross-Pitaevskii equation~\cite{Pitaevskii2003}; in a non-dimensional form it is given by:
\begin{equation}
	\label{eqn-gpe}
	\frac{d \psi}{dt} = -\frac{i}{2}\left(\nabla^2 - \left| \psi \right|^2\right) \psi,
\end{equation}
where in these units the quantized circulation around a single vortex line is given by: $\Gamma = \oint d\V\ell \cdot \V u = 2 \pi$.
The GPE has a characteristic length-scale, known as the `healing length', $\xi$, which corresponds to the size of the density-depleted region around each vortex core ($\xi = 1$ in our non-dimensional units if the background density is $\rho_0=1$). 
The GPE is a useful model system for studying vortex dynamics: vortex lines are  easily identified, reconnections occur without divergences in physical quantities, and the topological dynamics were recently shown to be comparable to real viscous fluids~\cite{Scheeler2014a}.  
Moreover, wavefunctions in the GPE can be numerically evolved using a simple split step method~\cite{Proment2012} (see supporting online material).

Due to the difficulty associated with generating initial states with vortices of arbitrary shape, previous studies of knotted superfluid vortices have been restricted to particular geometries within one knot family: the torus knots~\cite{Proment2012}.
Here, we directly integrate the flow field of a classical fluid vortex to produce phase fields with defects (vortices) of any topology or geometry~\cite{Scheeler2014a} (\figref[E]{fig-ideal}).
Using this technique, we are able to observe the evolution of all prime knot and links with 9 or less crossings, a total of 322 distinct topologies.

To ensure consistency between different topologies, we choose the `ideal' form of each knot, equivalent to the shape of the shortest knot tied in a finite thickness rope (\figref[B-D]{fig-ideal})~\cite{PPieranski1998,zotero-1830189-532}.
These canonical shapes are known to capture  aspects of the knot type as well as approximating the average properties of random knots~\cite{Katritch1996}.
To test the sensitivity of the evolution to scale and perturbations, we  simulate the dynamics at three different scales: $r_0/\xi = \{15, 25, 50\}$, where $r_0$ is the characteristic radius of loops in the ideal knots~(\figref[C]{fig-ideal}). 
We also consider four randomly perturbed copies of each $n \le 8$ knot/link ($4 \times 123$ configurations with r.m.s.~deviation $\sigma = 0.25 r_0$ and $r_0=15\xi$).
We label topologies using a generalized notation following \cite{zotero-1830189-531}, e.g.~a ``stevedore's knot" is K6-1, with the `K' indicating it is a knot (vs.~a link, `L'), $n=6$ is the minimal crossing number (\figref[F]{fig-ideal}), and the remainder indicates an arbitrary ordering. 

\begin{figure*}
\includegraphics{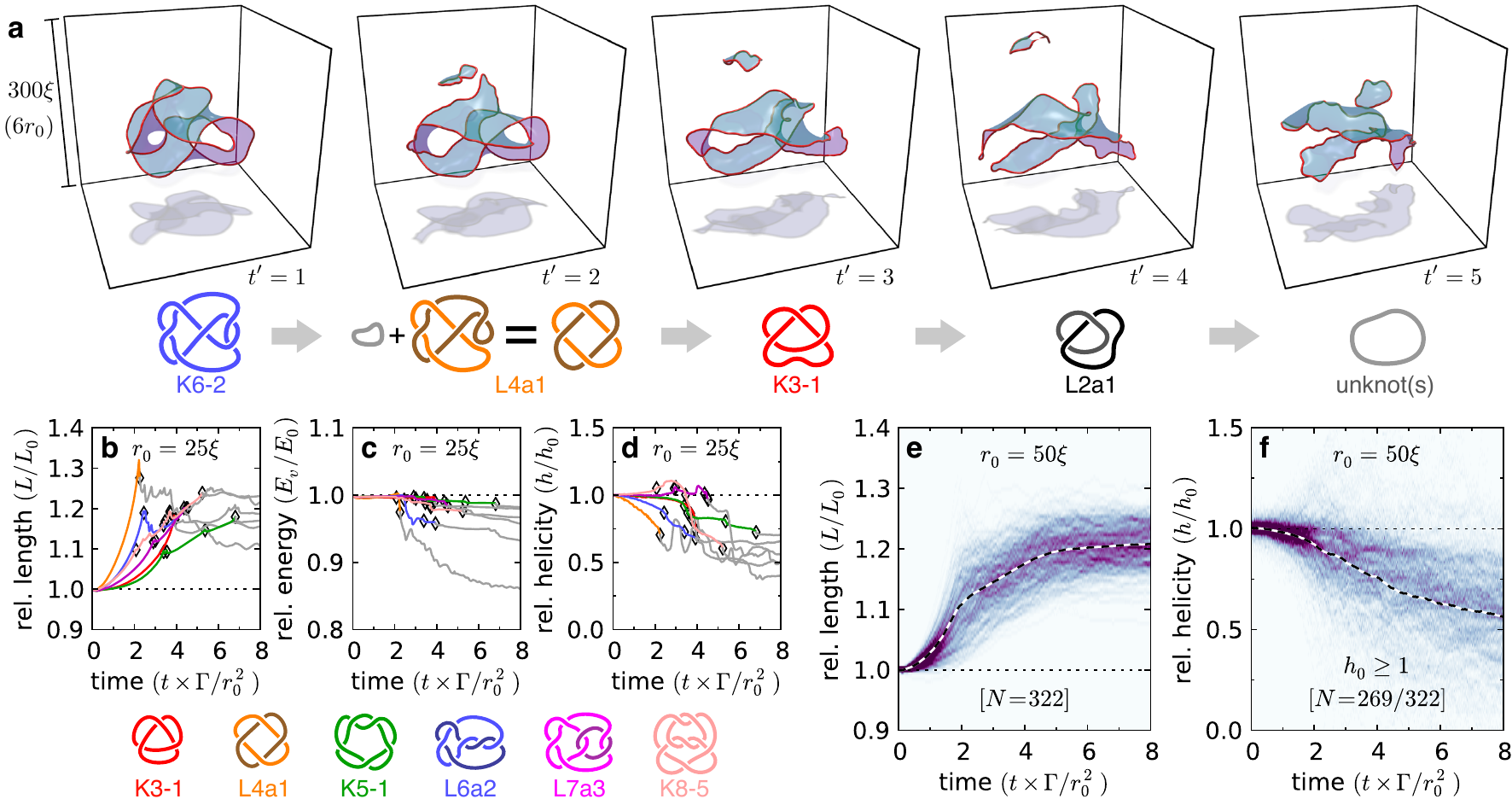}	
\caption{
	\label{fig-geo}
	Geometric evolution of vortex knots.
	\pp{A} The untying of a randomly distorted 6-crossing knot (K6-2, $r_0 = 50 \xi$) to a collection of unknotted rings.
	The rescaled time, $t' = t \times \Gamma / r_0^2$, is shown for each step.
	The top section shows density iso-surfaces of the wavefunction (red, $|\psi|^2 = 1/2$) and the transparent surfaces (teal or purple) show a constant phase iso-surface. 
	Each volume has been centered on the vortex, which would otherwise have a net vertical motion; only 48\% of the simulation volume is shown.
	% The topology of each time step is indicated below the volume.
	\pp{B-D} The relative length, vortex energy, and helicity as a function of time for six different ideal links/knots with $r_0 = 25 \xi$.
	% A diagram of each color-coded topology is shown below.
	Each topological jump is marked with a diamond, and the line color changes to gray after the knot has untied.
	\pp{E-F} 2D histograms of relative length and helicity as a function of time for all prime topologies with $n \le 9$.
	The dashed lines indicate average values.
	The helicity histogram, \pp{E}, only includes the 269/322 topologies with $h_0 \ge 1$.
}
\end{figure*}

\figureref[A]{fig-geo} shows the evolution of a 6-crossing knot, \hbox{K6-2}, as it unties.
The knot can be seen to deform towards a series of vortex reconnections that progressively simplify the knot until only unknotted rings (unknots) remain. 
This behavior has previously been observed for a handful of simple knots and links, here we find the same behavior in every of the 1,458 simulated vortex knots.
Significantly, this unknotting proceeds regardless of scale or distortions, indicating that it is a generic phenomenon. 

To quantify the dynamics of vortex knots as they untie, we compute their length, energy, and centerline helicity as a function of the evolving geometry of the vortex lines.
The non-dimensional `centerline helicity' -- which measures the total linking, knotting, and coiling in the field -- is given by~\cite{Moffatt1969,Berger1999,Scheeler2014a,Akhmetev1992}:
\begin{equation}
	\label{eqn-helicity}
	h = \sum_{i\neq j} Lk_{ij} + \sum_{i} Wr_i,
\end{equation}
where $Lk_{ij}$ is the linking number between vortex lines $i$ and $j$, and $Wr_i$ is the 3D writhe of line $i$, which includes contributions from knotting as well as helical coils.
Similarly, the vortex energy, $E_v$, can be computed purely in terms of the vortex geometry, up to a logarithmic core-correction factor~\cite{Donnelly2009}.
Note that this energy is not the total energy of the superfluid, which would also include sound waves and is conserved in the GPE.

\Figureref[E-F]{fig-geo} shows the length and helicity of all 322 topologies as they untie (see also \fref{S1} in the supporting materials).
Three general trends can be clearly discerned from our results: 
1) the timescale for unknotting  is determined predominantly by the overall scale of the knot,
2) the helicity is not simply dissipated, but rather converted from links and knots into helical coils, with an efficiency that depends on scale, and
3) the vortex lines stretch by $\sim$20\% as they untie, even though the vortex energy decreases slightly.
Interestingly, all of these results appear to be independent of complexity: simple knots untie just as quickly as complicated ones, and lose the same relative amount of helicity and energy (\fref{S2}).

A histogram of the unknotting times, \figref[A-D]{fig-histo}, is consistent with a log-normal distribution.
We find that once the time is appropriately rescaled, the mean unknotting times for each simulation group are in the range: $\left<t_{unknot}\right> \approx (3.5-4.0)\ r_0^2/\Gamma$ ($r_0$ is the diameter of the `rope' in which the ideal knot is tied). 

Despite the fact that the vortex knots untie, they do not lose all of their initial centerline helicity.
After each reconnection event, helices with a range of length scales are produced on the reconnected vortices.
Without any spatial cutoff, this process is expected to exactly conserve helicity~\cite{Scheeler2014a,Laing2015}, however, small helices (compared to the healing length) are radiated away as sounds waves.
As a result, we observe an average helicity loss which has an empirical $\Delta h / h_0 \propto (r_0/\xi)^{-0.5}$ trend, consistent with previous observations of trefoil knots~\cite{Scheeler2014a}. 

In all cases, the total vortex length increases until the knot is completely untied, at which point it approximately stabilizes. 
By contrast, the vortex energy is nearly constant except during reconnections, which dissipate a small amount.
As with helicity, the relative energy loss  decreases with increasing knot scale, however, it follows an approximate $\Delta E_v / E_0 \propto (r_0/\xi)^{-1.0}$ trend and so is not proportional to the loss in helicity.

If one assumes that concentrated vorticity distributions will always expand, these results have an intuitive explanation.
Collections of unknotted vortex rings may separate without stretching individual vortex lines, but this is not possible for a linked or knotted configuration.
At the same time, if the system is undriven the vortex lines must change configuration to conserve energy as they increase in length: as previously observed for simple knots, the formation of closely spaced, anti-parallel vortex pairs reduces the energy per unit length~\cite{Kleckner2013}.
As the stretching continues, these anti-parallel vortex lines are driven closer together until they ultimately reconnect; this process continues until the knots are completely untied.
Interestingly, such a picture also naturally produces the anti-parallel reconnection geometry that favors helicity conservation.

\begin{figure*}
\includegraphics{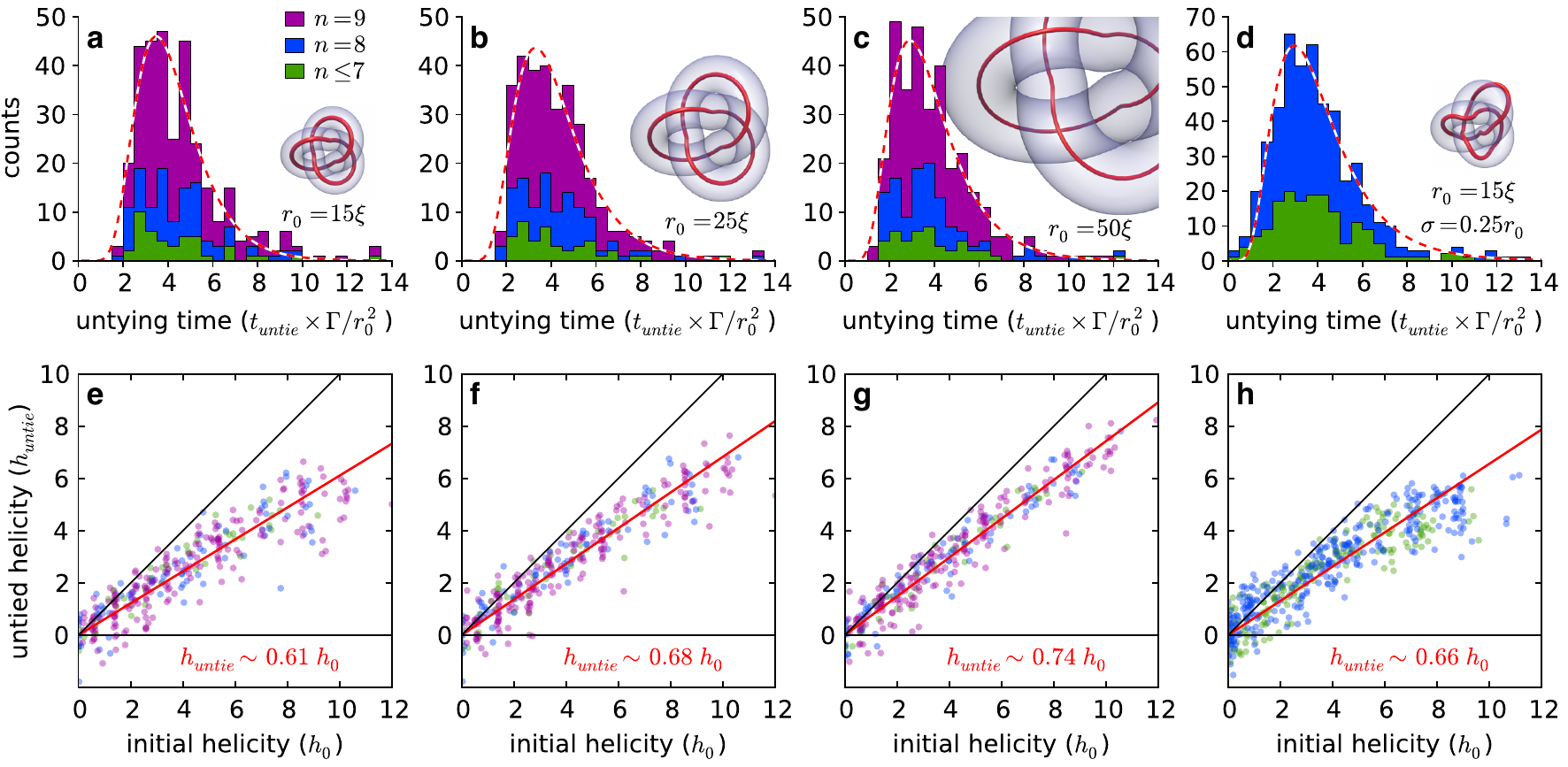}	
\caption{
	\label{fig-histo}
	Histograms of the rescaled untying time \pp{A-D} and the untied vs.~initial helicity \pp{E-H} for four different groups of simulations: \pp{A-C, E-G} all 322 ideal knots with $n\le 9$ at a scale of $r_0 = \{15, 25, 50\}\xi$ and \pp{D,H} four randomly distorted versions of each $n\le 8$ ideal knot with $r_0 = 15\xi$ and $\sigma = 0.25 r_0$ (492 simulations total).
	\pp{A-D} The distribution of untying times is well described by a log-normal distribution (dashed red line): $P(t) \propto \frac{1}{x} \exp\left[-\frac{(\ln x - \mu)^2}{2 \sigma^2}\right]$, where $\exp{\mu} = \{4.0, 3.9, 3.5, 3.7\}$ and $\sigma = \{0.37, 0.41, 0.44, 0.47\}$ for \pp{A-D}, respectively.
	\pp{E-H} The final helicity is approximately proportional to the initial helicity (red line).
	The degree to which helicity is preserved depends on overall scale, but is not slightly affected by randomly distorting the knots.
}
\end{figure*}

Although the above results demonstrate the overwhelming tendency for vortex knots to untie, they do not elucidate the specific topological pathways which produce this untangling.
To measure these unknotting sequences, we identify the topology, $T_i$, of the vortices after each reconnection by computing their HOMFLY-PT polynomials~\cite{Freyd1985,Przytycki1987} (see supporting materials).
This process reduces each simulation to an ordered list of visited knot types (e.g.~\figref[A]{fig-geo}), allowing us to consider the decay process in terms of topological invariants.
Due to the high degree of symmetry of ideal knots, reconnections are often nearly coincident in time, preventing identification of the intermediate topology.
To avoid this complication, we only consider the decays of the randomly distorted knots, which break this symmetry.

The first question we examine is  whether the knot is simplifying at each step.
We quantify the knot complexity via the crossing number, $n$, of each knot in a minimal 2D diagram (\figref[F]{fig-ideal}).
\Tableref{table-jumps} shows the statistics of the jumps in the crossing number through all reconnections, revealing knots are about an order of magnitude more likely to `untie' ($\Delta n < 0$) than `retie' ($\Delta n > 0$) at each individual reconnection. 
On average, more than one crossing is removed with each reconnection, underscoring the fact that physical reconnections of vortices in 3D are not equivalent to  removing (or  adding) a single crossing from a 2D minimal knot diagram.  
Nonetheless, the minimal diagrams reveal a clear trend towards  topological simplification.

\begin{figure*}
\includegraphics{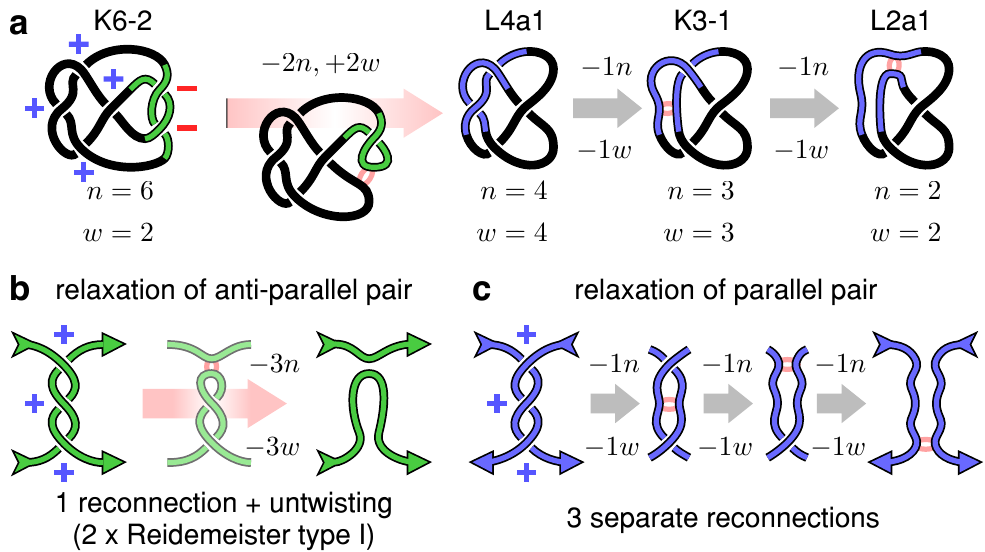}	
\caption{
	\label{fig-mechanisms}
	Topological mechanisms for untying vortex knots.
	\pp{A} Knot diagrams of an observed decay pathway for a 6 crossing knot; all steps can be described by local untwisting events.
	The minimal crossing number, $n$, and topological writhe, $w$, is labeled for each diagram.
	\pp{B, C} Nearly all reconnection events can be classified either as relaxation of a twisted pair in an anti-parallel or parallel orientation; in either case $\Delta n = -|\Delta w|$.  Reconnections of anti-parallel pairs are equivalent to a crossing removal plus one or more type-I Reidemeister moves.
}
\end{figure*}

\setlength{\tabcolsep}{0.2em}
\begin{table}
\begin{tabular}{p{1.7cm}p{0.9cm} p{1.7cm}p{0.9cm} p{1.7cm}p{0.9cm}}
	\hline
 	\ra$\Delta n  =  -1$:&\ra 34.4\% & %
 	\ra $\Delta n  =  -2$:&\ra 41.8\% & %
	\ra $\Delta n \le -3$:&\ra 12.5\% \\
	\ra $\Delta n  =  +1$:&\ra  3.1\% & %
	\ra $\Delta n  =  +2$:&\ra  5.8\% & %
	\ra $\Delta n \ge +3$:&\ra  0.3\% \\
	\hline
	\multicolumn{5}{p{7.5cm}}{\ra Removes crossings of same sign ($|\Delta n| = |\Delta w|$):} &\ra 96.1\% \\
	\multicolumn{5}{p{7.5cm}}{\ra Jump ends in maximally chiral ($|w_{final}| = n_{final}$):} &\ra 82.6\% \\
	\multicolumn{5}{p{7.5cm}}{\ra Jump leaves maximally chiral:} &\ra 0.7\% \\ 
	\hline
\end{tabular}
\caption{
	\label{table-jumps}
	Probabilities of topological jumps of various types.  
}
\end{table}

\begin{figure*}
\includegraphics{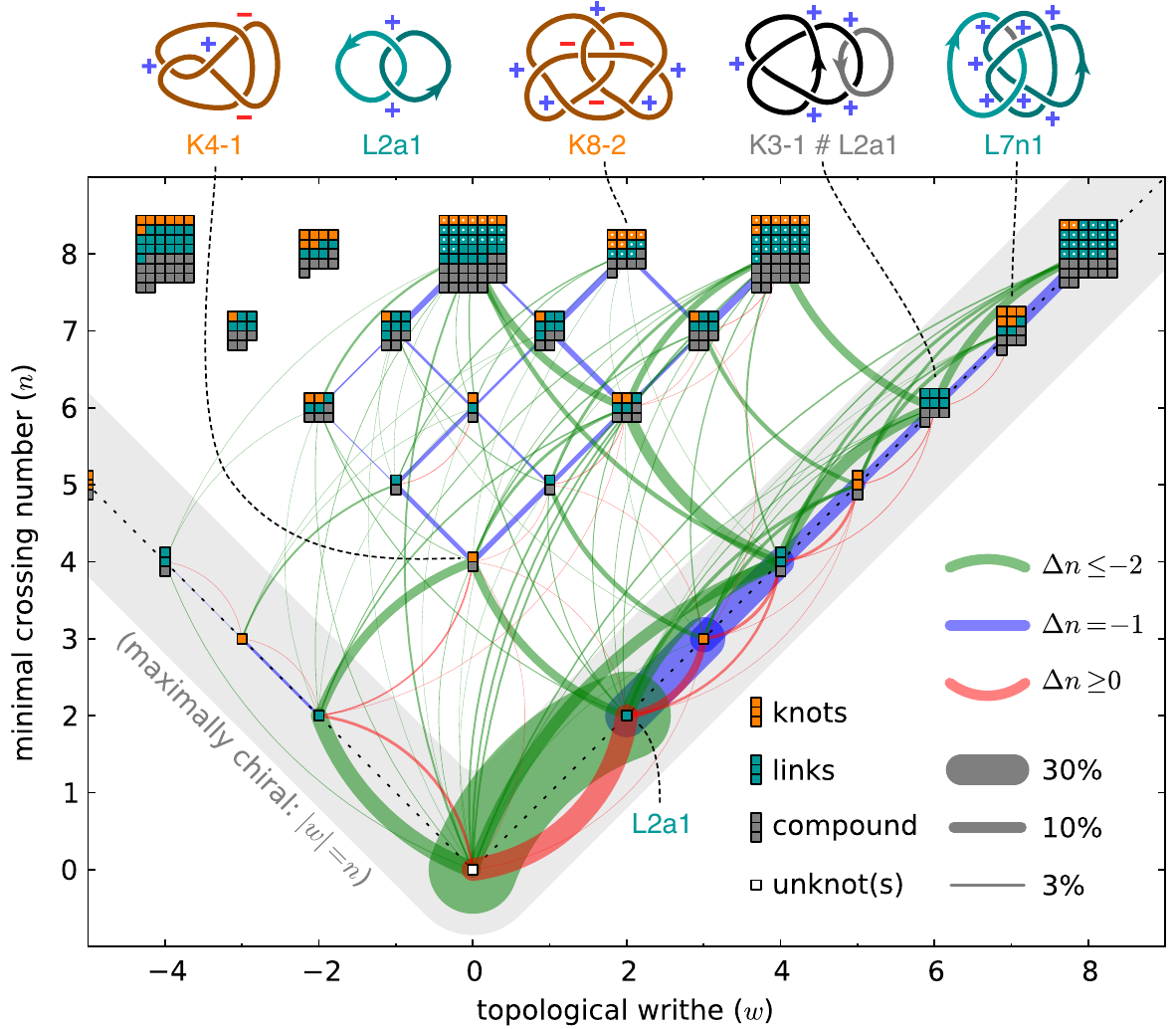}
\caption{
	\label{fig-pathways}
	Unknotting pathways in terms of minimal crossing number, $n$, and topological writhe, $w$.
	The squares indicate the total number of topologies at each point, including non-prime links/knots; five example topologies are shown on top.
	Four randomly distorted copies of each $n=8$ prime link/knot are used as the starting points (marked with white dots; only one handedness of chiral topologies is considered).
	 Green and blue lines indicate reconnections which reduce the crossing  number, while red lines show events which increase crossing number.
	 The lightly shaded region indicates the maximally chiral topologies.
}
\end{figure*}

If each reconnection does not correspond to `removing' a single crossing from a 2D knot diagram, is it still possible to produce an intuitive description of these events in terms of such diagrams?
This question can be answered by considering the 2D topological writhe, $w(T_i)$, which is obtained by summing the handedness ($\pm 1$) of each crossing in a minimal knot diagrams (obtained from~\cite{Chaa, Cha}).
Remarkably, we find that the vast majority (96.1\%) of reconnection events only remove crossings of the same sign from 2D diagrams, i.e.~$|\Delta n| = |\Delta w|$.
Such reconnections have a simple interpretation: they are equivalent to the relaxation of a parallel or anti-parallel pair in a 2D diagram (\figref{fig-mechanisms}).
The removal of multiple crossings thus occurs by a single reconnection happening in an anti-parallel pair, followed by the untwisting of a topologically trivial loop by type-I Reidemeister moves~\cite{Reidemeister1927,Alexander1926}.
Reconnections followed by more complicated simplifications are possible, however such events are observed to be very rare.

\figureref{fig-pathways} shows the topological writhe and crossing number of every knot with $n \le 8$ (including non-prime topologies), connected by lines indicated the frequency of the observed unknotting pathways.
In addition to illustrating the above results, this diagram reveals the importance of the `maximally chiral' topologies, for which $|w| = n$.
The topological writhe for any particular knot or link is bounded by the number of crossings; maximally chiral knots and links saturate this bound, which corresponds to every crossing having the same sign. 

Despite the fact that only around a third of all $n\le8$ topologies are maximally chiral, 82.6\% of jumps end in such a state.
The dominance of this pathway has a simple interpretation: if we assume all reconnections satisfy $|\Delta n| \ge |\Delta w|$ (corresponding to a slope of $|\Delta n / \Delta w| \ge 1$ in \figref{fig-pathways})~\footnote{
	Although the observation that $|\Delta n| \ge |\Delta w|$ seems self-evident when considering minimal crossing diagrams, we are not aware of a proof of this relationship.  
	Nonetheless, we never observe reconnections which violate it.
}, once the vortex knot decays into a maximally chiral topology it can only leave such a state by \emph{increasing} its crossing number.
Indeed, due to the `gap' between maximally and non-maximally chiral states, the crossing number must increase by $\Delta n \ge +2$ to leave the maximally chiral branch.
Moreover, even in the event that the crossing number does increase by this amount, we observe that it still typically stays on the maximally chiral branch.
Thus, statistically, most knots are funneled into maximally chiral pathway during their untying, after which they decay only along this pathway.

Our observation of a preferred maximally chiral pathway is a generalization of a previously known result for site-specific recombination of DNA knots:
any $p=2$ torus knot/link (which are all maximally chiral) may only convert into another $p=2$ torus knot via reconnections if the crossing number is decreasing~\cite{Shimokawa2013}.
Our results indicate that this torus knot pathway is one example of a more general phenomena.
Intuitively, this suggests untangling knots tend to end up in states which are twisted in only one chiral direction.

Taken as a whole, we find that the topological behavior of superfluid vortex knots -- even complicated tangles -- can be understood via simple principles.
All vortex knots untie, and they tend to do so efficiently: monotonically decreasing their crossing number until they are a collection of unknotted vortices.
This suggests that non-trivial vortex topology in superfluids -- or any fluid with similar topological dynamics -- should only arise from external driving.
Even in the presence of driving, the observed decay pathways indicate that vortices would likely settle into a maximally-chiral topology; it would be of great interest to probe for such states in superfluid or classical turbulence.

Our results relate the global geometry of vortices and their topology, and thus should be independent of the small-scale details of superfluid vortex reconnections.
Fundamentally, analogous reconnection events determine the evolution of topology in a wide range of fields, resulting in the already noted connections to classical viscous fluids and DNA.
The geometric and topological nature of our results suggests that they might apply more generally, forming a universal set of mechanisms for understanding the role knots play in a variety of physical systems.

\section{Acknowledgements}
The authors acknowledge M. Scheeler and D. Proment for useful discussions.
This work was supported by the National Science Foundation (NSF) Faculty Early Career Development (CAREER) Program (DMR-1351506), and completed in part with resources provided by the University of Chicago Research Computing Center and the NVIDIA Corporation.
W.T.M.I. further acknowledges support from the A.P. Sloan Foundation through a Sloan fellowship, and the Packard Foundation through a Packard fellowship.

\bibliography{refs}

\end{document}